\documentclass[prd, twocolumn, floatfix, nofootinbib, notitlepage, showkeys, showpacs]{revtex4-1}

\usepackage[mathscr]{euscript}
\usepackage{amsmath}
\usepackage{graphicx}
\usepackage{dcolumn}
\usepackage{bm}
\usepackage{epsfig}
\usepackage{amssymb,latexsym,mathrsfs}
\usepackage{graphicx}
\usepackage{color}
\usepackage{hyperref}
\usepackage{float}
\usepackage{diagbox}

\usepackage{tikz}
\usepackage[compat=1.1.0]{tikz-feynman}

\usepackage{amsmath}

\hypersetup{
    colorlinks=true,
    linkcolor=red,
    citecolor=blue,
}

\usepackage{amsmath}
\usepackage{amssymb}
\usepackage{subfigure}
\usepackage{hyperref}
\usepackage{url}
\usepackage{xcolor}
\usepackage{color}
\definecolor{amaranth}{rgb}{0.9, 0.17, 0.31}
\definecolor{purple(munsell)}{rgb}{0.62, 0.0, 0.77}
\definecolor{americanrose}{rgb}{1.0, 0.01, 0.24}
\definecolor{palatinateblue}{rgb}{0.15, 0.23, 0.89}
\definecolor{royalblue(web)}{rgb}{0.25, 0.41, 0.88}
\definecolor{hanpurple}{rgb}{0.32, 0.09, 0.98}
\definecolor{beaublue}{rgb}{0.74, 0.83, 0.9}
\definecolor{carminered}{rgb}{1.0, 0.0, 0.22}
\definecolor{brightpink}{rgb}{1.0, 0.0, 0.5}
\definecolor{vividviolet}{rgb}{0.62, 0.0, 1.0}

\definecolor{electron}{rgb}{1.0, 0.67, 0.22}
\hypersetup{ linktoc=all,
    colorlinks, linkcolor={palatinateblue},
    citecolor={brightpink}, urlcolor={amaranth}}

\newcommand{\be}{\begin{equation}}
\newcommand{\ee}{\end{equation}}
\newcommand{\bs}{\begin{split}} 
\newcommand{\bea}{\begin{eqnarray}}
\newcommand{\eea}{\end{eqnarray}}

\newcommand{\bes}{\begin{subequations}}
\newcommand{\ees}{\end{subequations}}

\newcommand{\bo}{\raise-1mm\hbox{\Large$\Box$}}

\begin{document}


\title{Extreme Electron Acceleration with Fixed Radiation Energy}

\author{Michael R. R. Good${}^{1,2}$}
\email{michael.good@nu.edu.kz}
\author{Chiranjeeb Singha${}^{3}$}
\email{chiranjeeb.singha@saha.ac.in}
\author{Vasilios Zarikas${}^{4}$}
\email{vzarikas@uth.gr}

\affiliation{
${}^1$ Energetic Cosmos Laboratory \& Physics Department, Nazarbayev University, Astana 010000, Kazakhstan\\
${}^2$ \quad Leung Center for Cosmology and Particle Astrophysics,
National Taiwan University, Taipei 10617, Taiwan\\
${}^3$ Theory Division, Saha Institute of Nuclear Physics, Kolkata 700064, India\\
${}^4$ \quad Department of Mathematics, University of Thessaly, Lamia 35100, Greece
}


\begin{abstract} 
We examine the extreme situation of radiation from an electron that is asymptotically accelerated to the speed of light, resulting in finite emission energy. The analytic solution explicitly demonstrates the difference between radiation power loss and kinetic power loss (null).
\end{abstract} 

\date{\today} 

\maketitle



\textit{Introduction. }-
Sending an electron to light speed usually results in a blast of infinite energy. Can this be avoided? Here, we demonstrate how to push an electron to its speed limit without a divergence in the emission energy. We compute the Larmor power, `Feynman power', angular distribution, and by analogy with the moving mirror model, the Bogoliubov spectrum.  The total energy radiated is computed and shown to be analytic and finite.  The disconnect between kinetic energy and radiation energy is sharp. There are three motivations for this work:
\begin{itemize}
    \item Black hole evaporation~\cite{Hawking:1974sw} leading to explosions~\cite{bhexplosions} have been unobserved thus far~\cite{Carr:2009jm}; insight into the process may be gained by studying asymptotic light speed acceleration radiation from non-exploding electrons.
    
    \item While moving point charge radiation is a staple of electrodynamics~\cite{Griffiths:1492149,Jackson:490457,Zangwill:1507229}, global, continuous, and dynamically accelerated rectilinear solutions are virtually absent in the literature.
    \item Although asymptotic inertia is a key requirement for unitarity, in the presence of a horizon it fails to conserve information~\cite{Good:2020uff}; studying an asymptotic inertial and horizon-free trajectory could provide insight into quantum purity. 
     
\end{itemize}

Furthermore, the lack of horizon formation coupled to asymptotic light speed is an interesting direction; see the early past of Carlitz and Willey~\cite{carlitz1987reflections}). A tractable global trajectory without a horizon and asymptotic light speed is rare.  Only two known solutions exist: (1) the `proex' trajectory~\cite{good2013time}, which is the limiting light speed case of the experimentally observed sublight speed drift of an electron emitting inner bremsstrahlung from beta decay~\cite{Good:2022eub}; and (2) the Light and Airy trajectory~\cite{Good:2021dkh}, which is the only known analytic solution for time evolution of particle creation from the quantum vacuum. The first solution, `proex', suffers from infinite emission energy if used to describe an electron in 3~+~1 dimensions~\cite{Good:2016yht}. For this reason, we focus on the second solution.  We are able to extend the 1~+~1 dimensional mirror model to the more realistic situation of an electron accelerating in 3~+~1 dimensions.

For clarity, we consider uniform acceleration; while very mathematically simple, it is not physically realistic as a global trajectory. For instance, a single electron emits an infinite amount of radiation energy when travelling eternally at uniform acceleration. Of course, while this makes sense because of the infinite energy required to accelerate it forever, it is easy to see this visually for the hyperbola of uniform acceleration because there are horizons, i.e., in null coordinates $u=t-x$ and $v=t~+~x$, a horizon is when $u(v)$ goes to infinity at some finite value of $v$ or $v(u)$ goes to infinity at some finite value of $u$. 

This long-persistent global definition problem associated with horizons has given rise to theoretical misunderstandings, and will continue to lead physical interpretation astray if regularization of the emitted radiation is not taken into account when global context matters. Infinite radiation energy plagues worldline examination of physical connections between kinetic power loss and radiation power loss. 

Collective attention should be paid to directly addressing radiation energy independent of kinetic energy considerations. As is already well-understood, non-uniform accelerated trajectories that drop to zero are perfectly capable of regularizing the global total emitted radiation energy. The trade-off is the lack of simplicity, and consequently, tractability (for instance, when determining the global radiation spectrum).    

We present a result in this article that provides a well-defined, analytic, simple, consistent, and complete solution for finite radiation energy and its corresponding spectrum. Uncommon solutions of this type can be immediately used to investigate questions of interest wherever globally continuous equations of motion are needed. For instance, the solution is well-suited for applications such as study via an Unruh--DeWitt detector trajectory~\cite{Cong:2020nec} or the non-uniform Davies--Fulling--Unruh effect~\cite{Doukas:2013noa}. 

Additionally, this work provides a simple conceptual and quantitative analog approach to understanding the radiation emitted by an electron that asymptotically advances to light speed, in a similar way as a uniformly accelerated electron asymptotically approaches light speed; however, the differences are crucial to analytic tractability: there is no horizon, the acceleration drops to zero, and the radiation energy is finite (though it nonetheless travels to light speed!). We emphasize that the analog approach is general enough to be applied to any rectilinear trajectory that emits finite radiation energy; however, it cannot be applied to uniform acceleration because of above aforementioned problems associated with the horizons. 

We analytically compute the relevant quantities for the specific extreme solution and demonstrate consistency between the various ways of computing the radiation energy.  The analog approach treats the electron as a tiny double-sided moving mirror; interestingly, this reveals previously unknown electron acceleration radiation spectra as well as a general connection between classical acceleration radiation and quantum acceleration radiation. 

The outline of the paper is as follows. First we provide context and consider the trajectory. Second, we compute the power and the self-force, examining the time-evolution. Third, we compute the `local acceleration', determine the conditions required for avoiding a divergence in the emission energy, and find the angular distribution of power. Fourth, we propose a candidate Bogoliubov spectrum, and lastly, we perform consistency checks and compute the total energy emitted.

\section{Infinite Emission Energy in the Causal Limit} 
Imagine one has an arbitrarily large (infinite) amount of energy to accelerate an electron, ignoring the mechanism that is doing the acceleration while assuming nonetheless that the electron accelerates along a specified sub-light speed trajectory in a time-like fashion to asymptotically approach the speed of light.  This would be the case, for instance, when attempting to push an electron along an eternal uniformly accelerated trajectory.  Such a motion would properly result in an infinite emission of radiation energy. 

This is the case when considering the scaling of radiated energy from trajectories that do not even have horizons. For instance, let us examine the case of emission energy associated with lowest-order inner bremsstrahlung~\cite{Jackson:490457} and the scaling with respect to the final electron speed $s$ (see~\cite{Good:2022eub}): 
\begin{equation}
E \sim \frac{1}{s}\ln \left(\frac{1~+~s}{1-s}\right)-2.\label{IB}
\end{equation}
When the charge approaches the speed of light, i.e., when $s\to 1$, then $E \to \infty$. A mathematical divergence in the emission energy occurs despite the lack of horizon. Of course, we must make it absolutely clear that an electron cannot be accelerated to the speed of light in the real world, due to limitations on the finite amount of available energy to give to the electron.  However, without even considering the nonphysical nature of the infinite amount of energy required to push a massive charged particle, it is physically clear from Equation~(\ref{IB}) that a radiating electron always travels at sub-light speeds (even asymptotically); otherwise, an infinite amount of radiation energy occurs in the expression for the emitted~energy.   

Thus, equipped with large amounts of available energy, could it be possible to push an electron to the speed of light without a divergence appearing in the total emission energy?

\section{Trajectory} 
\begin{figure}[H]
  \includegraphics[width=0.95\linewidth]{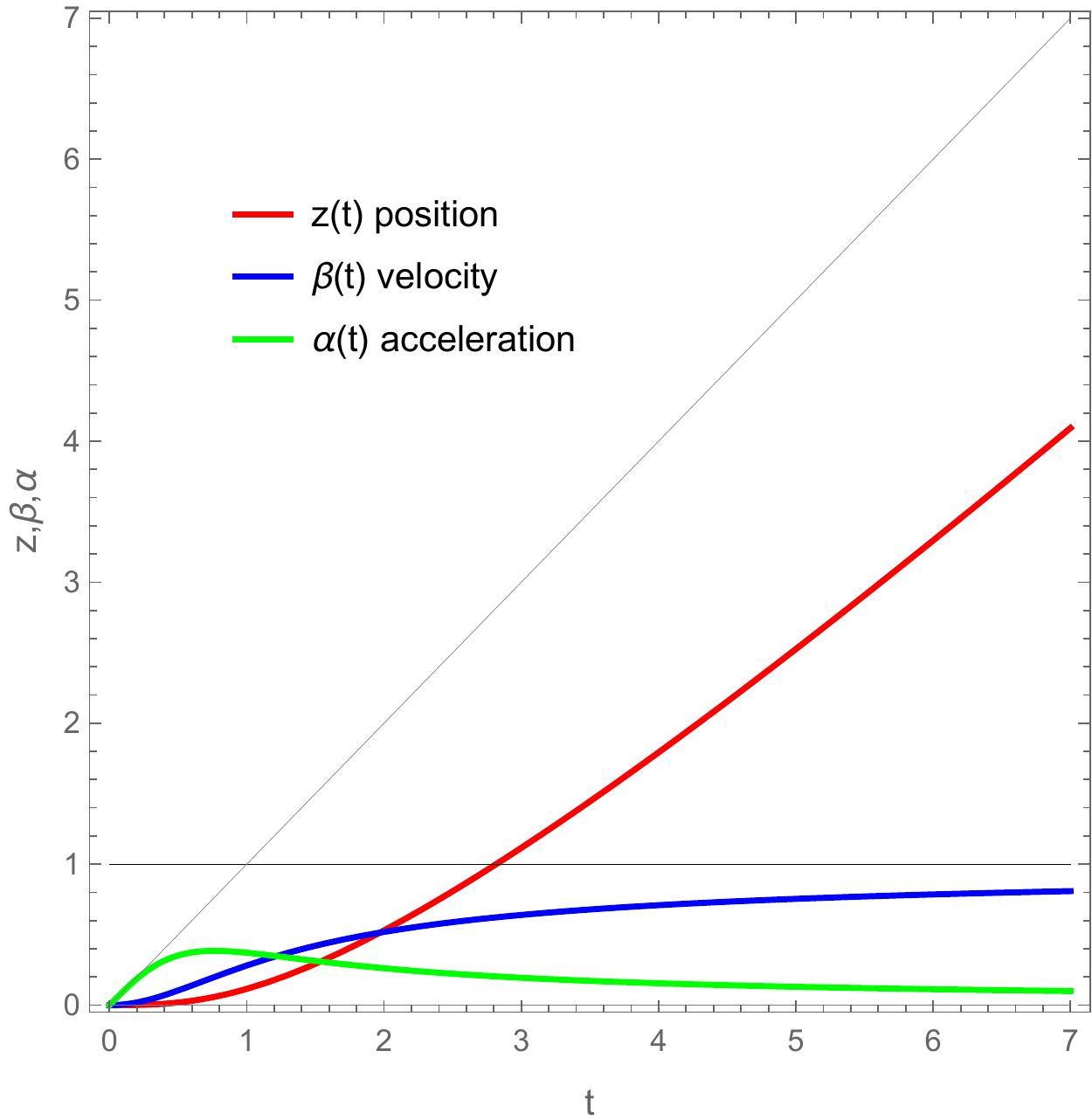}
 \caption{The position $z(t)$, velocity $\beta(t)$, and proper acceleration $\alpha(t)$ plotted to demonstrate the position (red) approaching a near 45 degree angle (diagonal null-ray is gray) as the velocity (blue) approaches the speed of light (black), while the acceleration (green) drops to zero.  Here, $\kappa =1$. The key takeaway is that an this motion has an asymptotic end state with drifting or coasting velocity at the speed of light, and is coupled to asymptotic zero acceleration. }
\label{three}
\end{figure}

Allow us to consider an electron following the light-cone advanced time $v=t~+~z$, trajectory~\cite{Good:2021dkh} 
\begin{equation}
v(u) = u ~+~ \frac{\kappa^2 }{3}u^3,\label{p(u)}
\end{equation}
with a single parameter $\kappa$ which sets the acceleration scale of the system.  Here, $u = t-z$ is retarded time. The electron moves in a straight-line path along one spatial direction (rectilinear), $z$, where
\begin{equation} z = \frac{v-u}{2} = \frac{\kappa^3 u^3}{6\kappa}, \quad \textrm{and,} \quad u = t-z = \frac{1}{\kappa}(6\kappa z)^{1/3}.\end{equation}
The rapidity $\eta(u)$, using $2\eta(u) \equiv  \ln v'(u)$, where the prime is a derivative with respect to the argument, is 
\begin{equation}
\eta(u) = \frac{1}{2} \ln \left(\kappa ^2 u^2~+~1\right).\label{eta(u)}
\end{equation}
From the rapidity, the velocity along the direction of motion is easily found to be $\beta \equiv \tanh \eta$. Plugging in Equation~(\ref{eta(u)}) provides
\begin{equation}    
\beta(u) = \frac{\kappa^2 u^2}{\kappa^2 u^2 ~+~ 2}\,, \label{beta(u)} 
\end{equation}  
and the proper acceleration Lorentz scalar, which follows from $\alpha(u) \equiv  e^{-\eta(u)} \eta'(u)$, is
\begin{equation}
\alpha(u) = \frac{\kappa ^2 u}{\left(\kappa ^2 u^2~+~1\right)^{3/2}}.
\end{equation}
The asymptotic approach to the speed of light is explicit: $\beta(u) \to 1$, as $u\to \pm \infty$.  Interestingly, the brakes are being applied, as $\alpha(u) \to 0$ as well.  Notice that $\alpha = \beta = 0$ at $u=0$.  The position $z(t)$, velocity $\beta(t)$, and the proper acceleration $\alpha(t)$ are plotted in Figure \ref{three} to illustrate the dynamical properties of the motion.  For further visual clarity, a parity-reversed plot of the trajectory from Equation~(\ref{p(u)}) is provided in a spacetime plot, and a (1~+~1) dimensional Penrose diagram is provided via Figures 1 and 2 of~\cite{Good:2021dkh}.

\section{Larmor power} 
\begin{figure}[H]
  \includegraphics[width=0.95\linewidth]{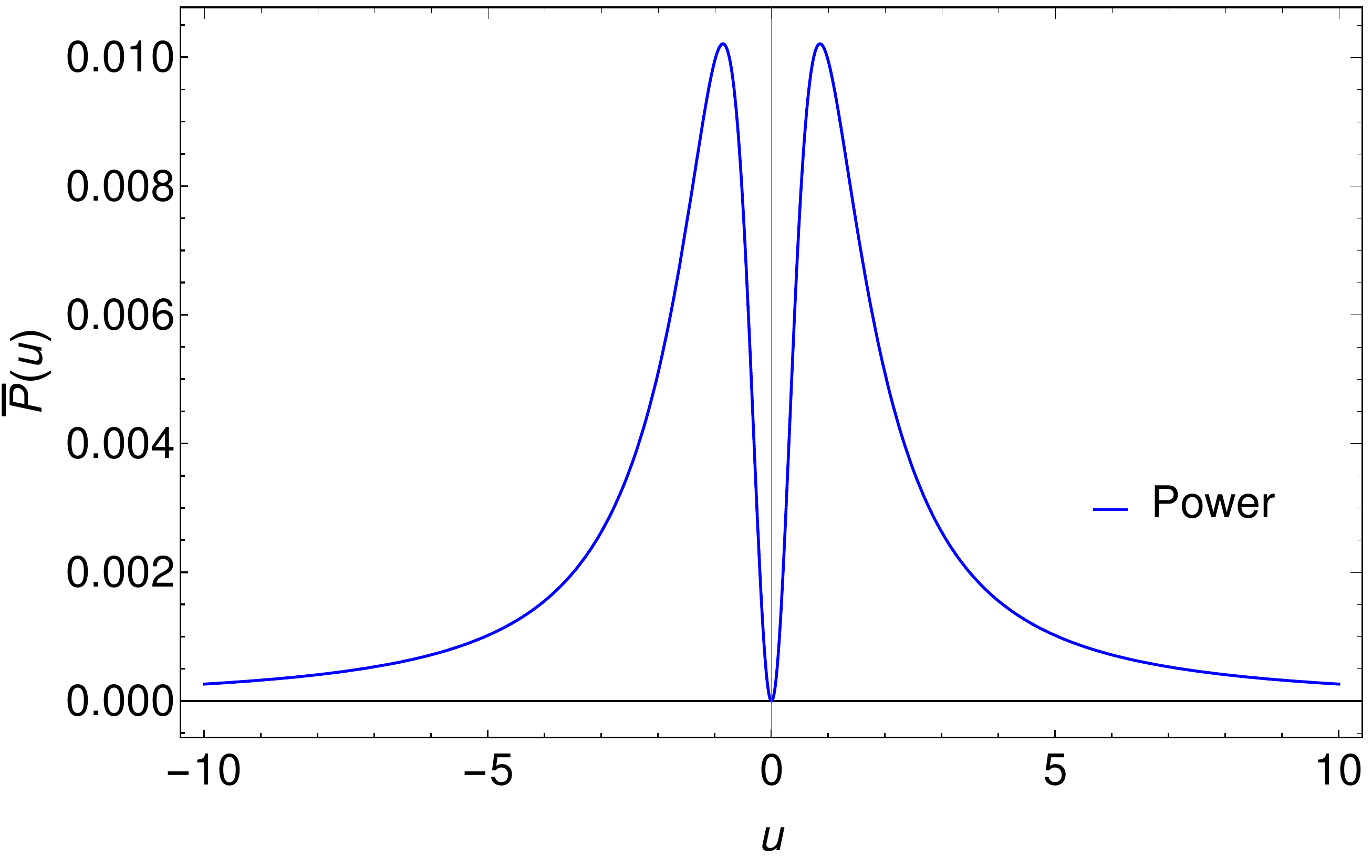}
 \caption{A plot of the Larmor power, $\bar{P}(u)$, from Equation~(\ref{barP}). From Equation~(\ref{barPF}), the integral under the curve is the energy,. Note that the Larmor power is always positive, asymptotically zero, and time-symmetric about $u=0$, where $\bar{P} = 0$.  Here, $\kappa =1$. }
\label{power}
\end{figure}

By working with unit charge and written as $\bar{P} = \frac{d E}{d u},$ such that
$\bar{P} = P \frac{d t}{d u} = P/(1-\beta)$, where $P = \alpha^2/6\pi$, we find the Larmor power $\bar{P}(u)$, formulated in terms of retarded time $u$, providing us with 
\begin{equation}
\bar{P} = \frac{\kappa ^4 u^2 \left(\kappa ^2 u^2~+~2\right)}{12 \pi  \left(\kappa ^2 u^2~+~1\right)^3}.\label{barP}
\end{equation}
Key takeaways from the Larmor power include the asymptotic end to the radiation at early and late times, symmetry in time, and zero emission at $u=0$.  Notice that the Larmor power emission is always positive.  This is in stark contrast to the negative energy flux commonly associated with asymptotic inertial mirrors in 1~+~1 dimensions from the quantum stress tensor energy flux, e.g.,~\cite{Walker:1984ya}.  
A plot of $\bar{P}(u)$  illustrates the power in Figure~\ref{power}.   We make special note that the Larmor formula describes radiation in the `far zone', and does not tell the story closer to the radiation charge.   For example, if a (classical) electric dipole starts to collapse at a certain time, the Larmor formula describes the radiation in the far zone; however, close to the dipole the energy flows from that initially stored in the electromagnetic field onto the collapsing dipole; see, e.g.,~\cite{hans}.

\section{Feynman power}
The Larmor power, $P=\alpha^2/6\pi$, or $\bar{P}$, Equation~(\ref{barP}), is not sufficient to understand the full story of the radiation. For instance, it is often supposed that the Larmor power tells us `when' the energy is radiated; however, this is not true for all motions. Larmor's formula can only be valid for certain types of motions, such as horizonless trajectories with finite global energy emission.  In fact, this insufficiency has led to considerable confusion when dealing with the equivalence principle and the question of whether an electron radiates while sitting in a gravitational field~\cite{Feynman:1996kb}.  

To better understand the radiated energy, it is beneficial to determine the force of radiation resistance and the work against this force, which is a measure of the energy loss due to radiation emission.  

\begin{figure}[H]
  \includegraphics[width=0.95\linewidth]{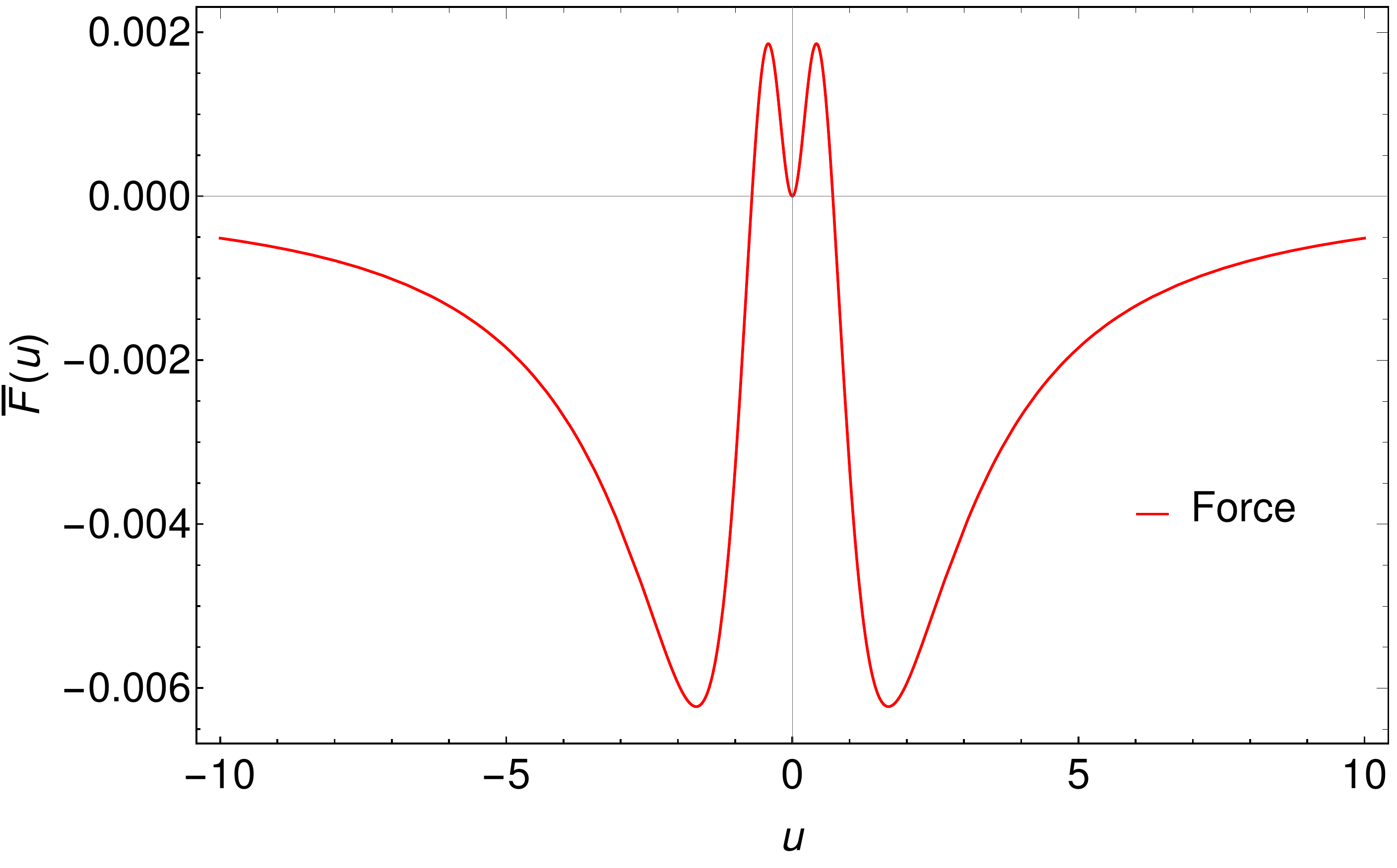}
 \caption{The above graph is of the Feynman power, or `self-force measure', from Equation~(\ref{barF}). Here, $\kappa = 1$.  The integral under the curve is the sign-flipped radiated energy, or the work done by the self-force, from Equation~(\ref{barPF}). Note that the Feynman power possesses negative dips, which contrasts with the always-positive Larmor power shown in Figure \ref{power}. }
\label{force}
\end{figure}

Therefore, we now turn to the self-force, $F=\alpha'(\tau)/6\pi$~\cite{Myrzakul:2021bgj,Ford:1982ct}; we find it convenient to dub this the `Feynman power'~\cite{Feynman:1996kb}. This is an analogous `self-force measure' $\bar{F}(u) = F\frac{d r}{d u} = F \beta/(1-\beta)$ as a function of retarded time $u$:
\begin{equation}
\bar{F} = \frac{\kappa ^4 u^2(1-2 \kappa ^2 u^2)}{12 \pi  \left(\kappa ^2 u^2~+~1\right)^3}. \label{barF}
\end{equation}
Unlike the Larmor power, the Feynman power can be negative.  Indeed, it must be negative for asymptotic inertial motions, as non-trivial changes in acceleration that asymptotically approach zero are characterized by sign flips (`what goes up must come down'). Similar to the Larmor power, the self-force and Feynman power are asymptotically zero at early and late times: $\bar{F} = 0$ at $u=0$. A plot of the Feynman power is shown in Figure \ref{force}. Notice the stark contrast between the Larmor power in Figure \ref{power} and the sometimes negative Feynman power in Figure \ref{force}, which obviously tell very different, yet globally consistent, stories about the power of the emission.

\section{Local acceleration}
Consider the so-called `local acceleration'~\cite{CW2lifetime}, $\bar{\kappa}(u)= \partial_u \ln v'(u) = v''/v'$, or in spacetime coordinates,
\begin{equation} \kappa(t) = \frac{2\ddot{z}}{(1-\dot{z})^2(1~+~\dot{z})}.\end{equation}
The local acceleration is a useful measure of the dynamical trajectory; in our case its value, as a function of retarded time $u$, is,
\begin{equation}
 \bar{\kappa}(u) = \frac{2 \kappa ^2 u}{\kappa ^2 u^2~+~1}. \label{locala}
 \end{equation}
The local acceleration is a particularly good quantity for gauging the thermality of the system, i.e., when it asymptotically approaches a constant $\bar{\kappa} \to \kappa$, the radiation is Planck-distributed with $T=\kappa/2\pi$. 

\begin{figure}[H]
  \includegraphics[width=0.95\linewidth]{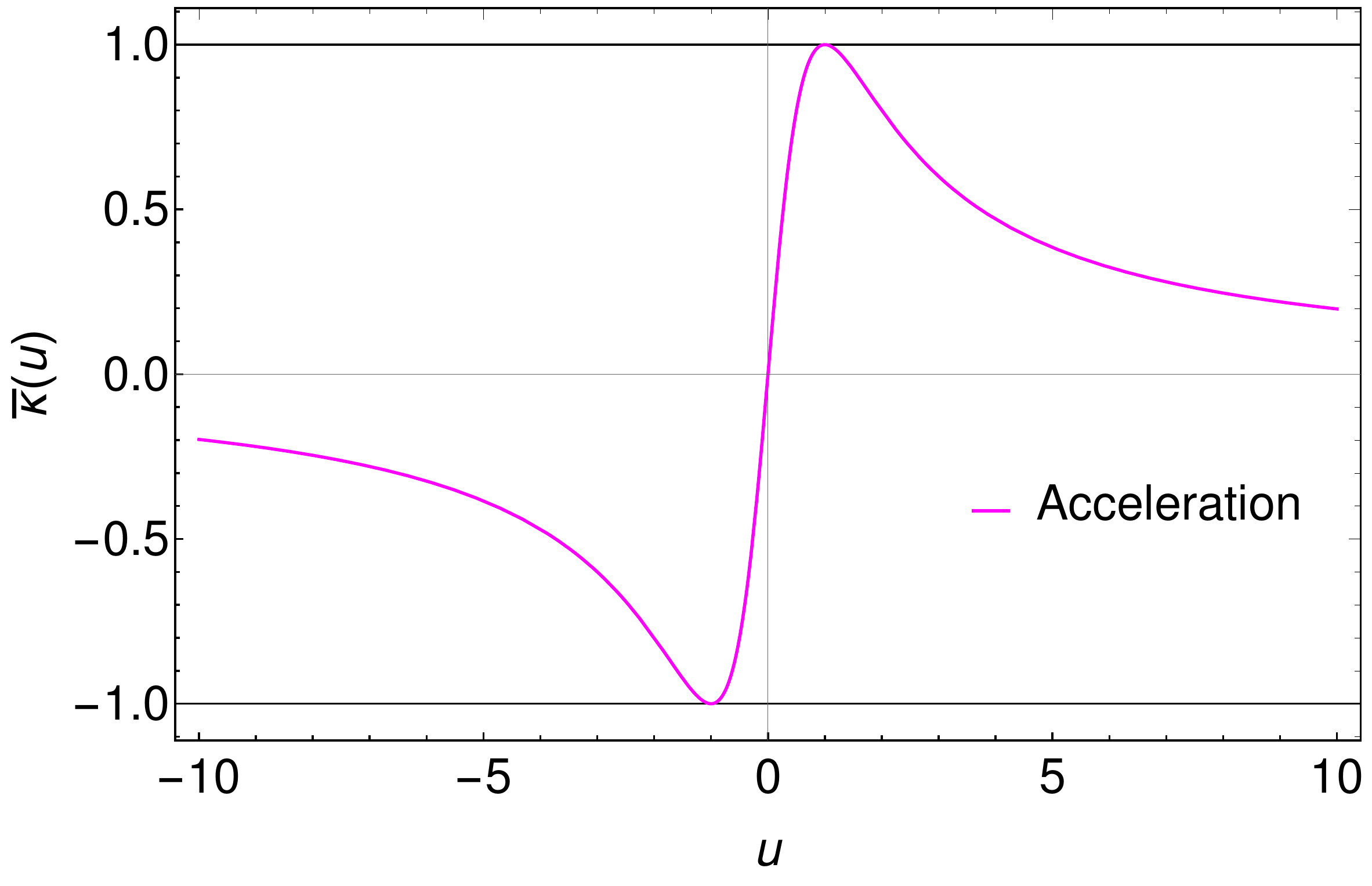}
 \caption{Plot of the `local' or `radiation' acceleration $\bar{\kappa}(u)$ from Equation~(\ref{locala}), which is a measure of the asymptotic inertia.  The asymptotic approach to zero insures no divergence in the radiated energy.  At $u=1/\kappa$, $\bar{\kappa}(u) = \kappa$.  Here, $\kappa = 1$.  Note the time asymmetry of the local acceleration, in contrast to Figures \ref{power} and \ref{force}.  }
\label{local}
\end{figure}

However, even though in our case $\bar{\kappa}(u) = \kappa$ when $u=1/\kappa$, our trajectory is non-thermal (non-Planckian), and thus the local acceleration can help us.  This provides a tool to understand how to avoid a divergence in the radiation energy.  Consider the analog between mirrors and electrons (namely, powers~\cite{Good:2021ffo,Zhakenuly:2021pfm} and self-forces~\cite{Myrzakul:2021bgj,Ford:1982ct}) outlined in the last four equations of Section \ref{sec8}. A connection between moving mirrors and moving charges has been established by Ritus~\cite{Ritus:2003wu,Ritus:2002rq,Ritus:1999eu,Nikishov:1995qs}, who found symmetries linking the creation of pairs of massless bosons or fermions by an accelerated mirror in 1~+~1 space and the emission of single photons or scalar quanta by electric or scalar charges in 3~+~1 space.   Walker~\cite{walker1985particle} has pointed out how the local acceleration $v''/v'$ must vanish, as it is the boundary term for an integration by parts when computing the total radiated energy. As can be seen, in Equation~(\ref{locala}) our $\bar{\kappa}(u)$ vanishes as $u\to\infty$. Thus, despite the asymptotic approach to light-speed $\beta  \to 1$ (in the forward direction toward the far away observer), the proper acceleration $\alpha \to 0$ dies off more quickly., that is, there is tension between the speed and the acceleration:
\begin{equation} 
\bar{\kappa}(u) \sim e^\eta \alpha = \sqrt{\frac{1~+~\beta}{1-\beta}} \alpha.
\end{equation}
The asymptotic inertia characterized by the acceleration must be strong enough to overcome the infinite rapidity. Figuratively, it is as if the observer finishes detection of the radiated energy before the electron smashes into the equipment. In order to avoid a divergence in the radiation energy, it is not enough to have asymptotic inertia when approaching to the speed of light; one must have extreme braking acceleration as well, and for the overall local acceleration one must have $v''/v' \to 0$.  A plot of the local acceleration is shown in Figure \ref{local}.  

\section{Distribution}
\begin{figure}[H]
  \includegraphics[width=0.8\linewidth]{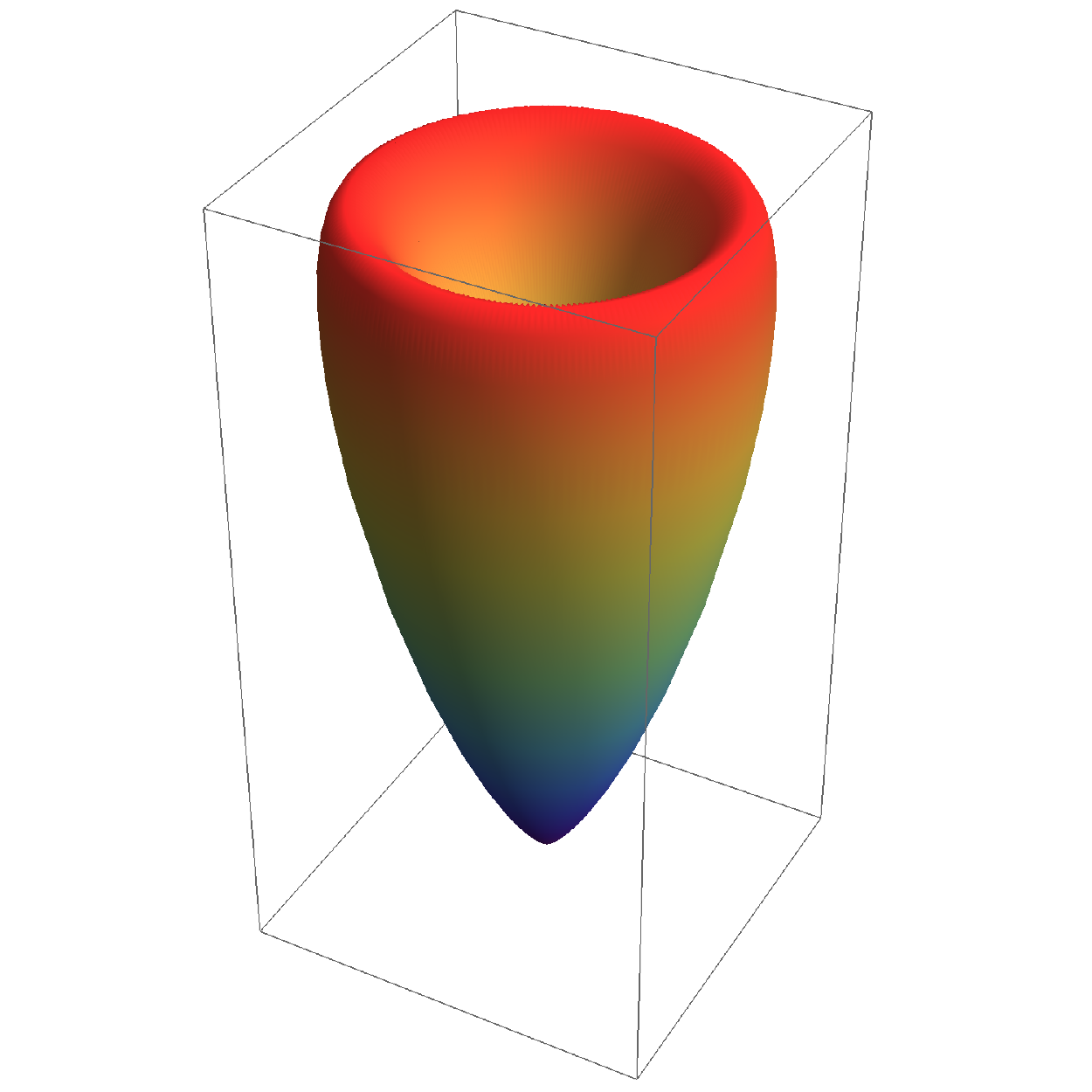}
 \caption{The power distribution in Equation~(\ref{shape1}) illustrated in a spherical plot. The shape, as expected for rectilinear braking radiation, is about the forward direction, not in the forward direction. Here, $\kappa =1$ and $\beta = 0.90$.}
\label{shape}
\end{figure}

To highlight the 3~+~1 dimensional character of the solution, we compute the power distribution for the charge in relativistic motion.  The power distribution has a well-known form (e.g., Jackson~\cite{Jackson:490457} or Griffiths~\cite{Griffiths:1492149}) for rectilinear motion.   We can find the distribution with straightforward vector algebra (see, e.g., the procedure in~\cite{Good:2019aqd} and the notation details in Appendix A.1), consistent with rectilinear motions,
\begin{equation} \frac{d \bar{P}}{d \Omega} = \frac{\kappa ^2 \beta (1-\beta)^4 \sin ^2\theta}{8 \pi ^2 (1-\beta  \cos \theta )^5},\label{shape1}
\end{equation}
where $\beta = \beta(u)$ is the time-dependent velocity from Equation~(\ref{beta(u)}). The $\hat{z}$ axis points along $\vec{\beta}$.
A spherical plot is shown in Figure \ref{shape}. Integration over solid angle $d \Omega = \sin\theta d \theta d \phi $ provides
\begin{equation}
 \bar{P} = \frac{\kappa ^2\beta (1-\beta)^4 }{3 \pi  \left(1-\beta ^2\right)^3},
 \end{equation}
which is identical to Equation~(\ref{barP}). Integration of the distribution from Equation~(\ref{shape1}) over time provides the time-independent distribution or angular differential distribution of the radiated energy:
\begin{equation} \frac{d E}{d \Omega} = \frac{5 \kappa }{512\sqrt{2}\pi}\frac{ \sin ^2\theta }{(1-\cos \theta)^{3/2}}.
\end{equation}
This can be integrated over the solid angle to find the total energy, resulting in Equation~(\ref{energy}). 

\section{Bogoliubov spectrum}\label{sec8}
To determine particle count and particle spectrum in a quantum context, it is worthwhile to compute the Bogoliubov coefficients in the style that Hawking~\cite{Hawking:1974sw} did for black hole particle creation.  In the same way that particle creation by black holes can be found with these coefficients, particle creation by a moving mirror can be found with the beta Bogoliubov coefficients~\cite{Birrell:1982ix}.  

\begin{figure}[H]
  \includegraphics[width=0.95\linewidth]{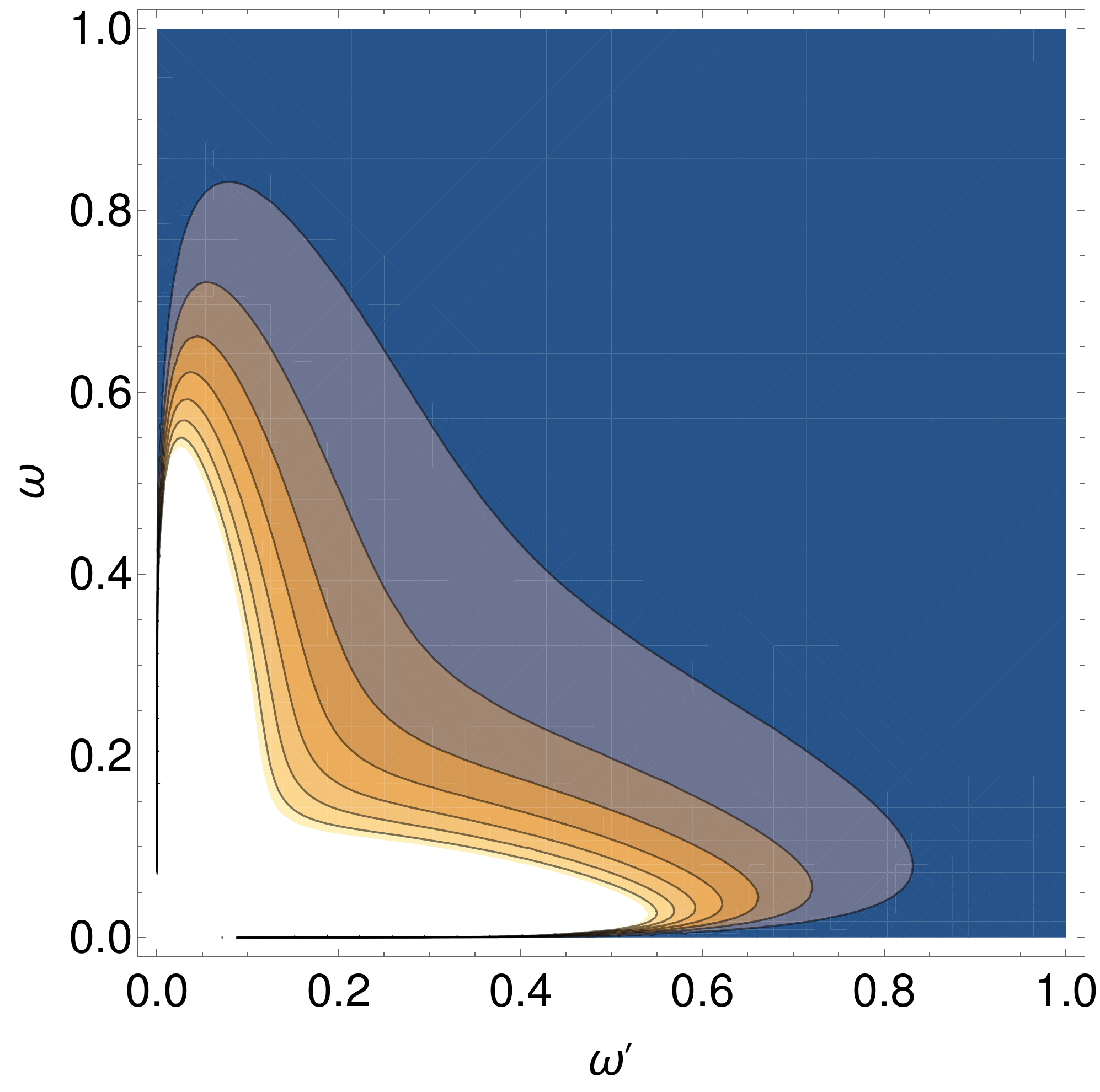}
 \caption{The $|\beta_{\omega\omega'}|^2$ spectrum of Equation~(\ref{betaT1}) in a contour plot.  The symmetry demonstrates that the summing of either $\omega$ quanta (Equation~(\ref{numerical1})) or $\omega'$ quanta (Equation~(\ref{energy})) results in the same energy being radiated,.}
\label{contour}
\end{figure}

\begin{figure}[H]
  \includegraphics[width=0.95\linewidth]{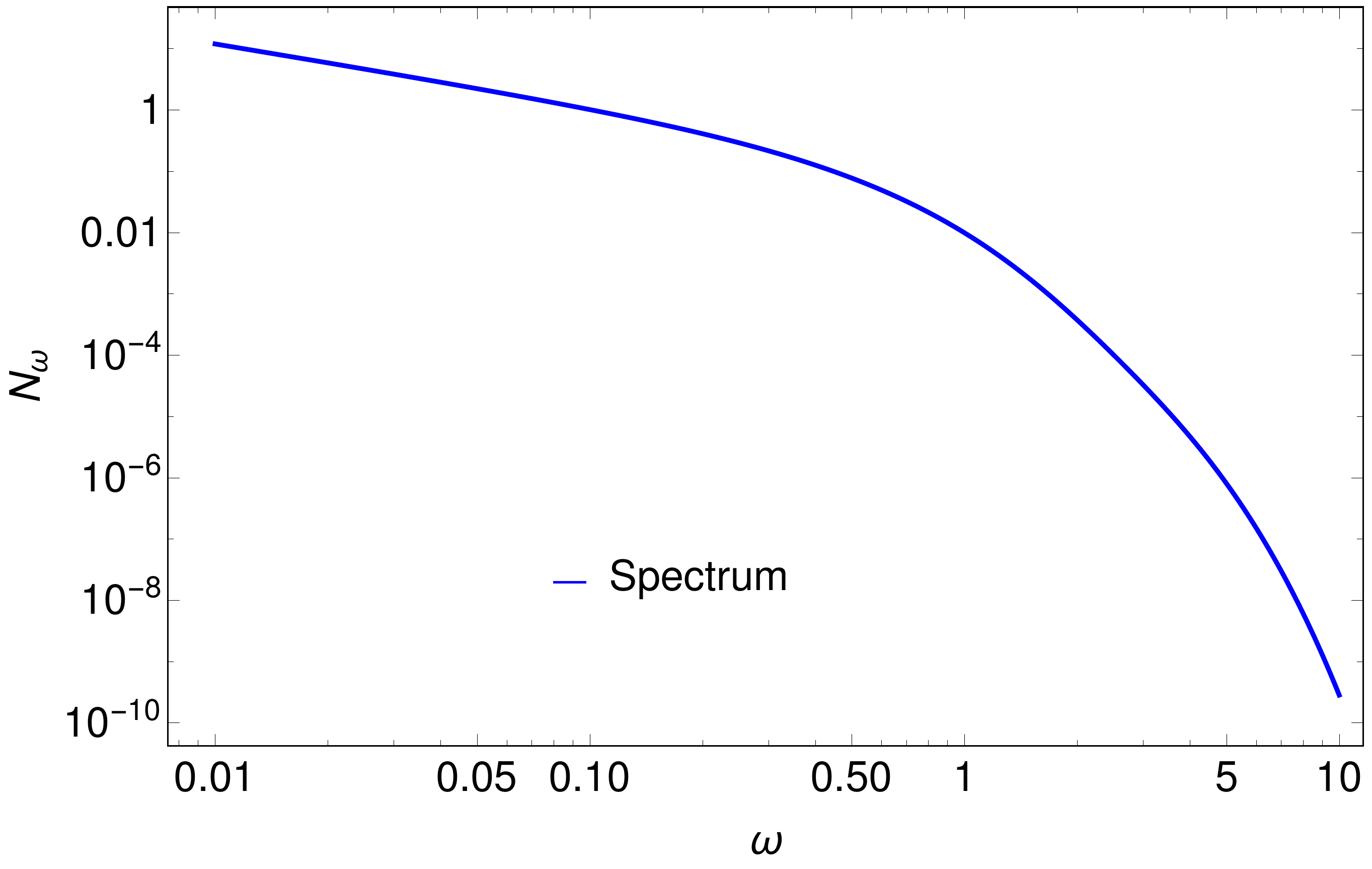}
 \caption{The spectrum $N_\omega$ (Equation~(\ref{Nw})), which is number of particles at any given frequency $\omega$, demonstrating the usual infrared divergence that signals a divergent total particle count $N$. Such a divergence is absent with asymptotic resting motions.}
\label{Number}
\end{figure}

Here, the beta Bogoliubov coefficients $\beta_{\omega\omega'}$, in the spirit of the analogy that a moving charge has with a moving mirror, are tractable for the spectrum dependent on the acceleration parameter $\kappa$. Here, $\omega'$ is the incoming frequency mode and $\omega$ is the outgoing frequency mode.  We use the same procedure and result of~\cite{Good:2021dkh}, combining the results for each side of the mirror in~\cite{Good:2021dkh}, both left and right, by adding the squares of the beta Bogoliubov coefficients. The beta coefficient can be found by
\begin{equation}
\beta_{\omega\omega'} = \frac{-1}{2\pi}\sqrt{\frac{\omega}{\omega'}}\int_{-\infty}^{~+~\infty}du\, e^{-i\omega u-i\omega'v(u)}\ .  \label{eq:bww} 
\end{equation} 
The overall spectrum is
\begin{equation} |\beta_{\omega\omega'}|^2 =\frac{\omega ' \text{Ai}\left(\frac{\omega ~+~\omega '}{\kappa ^{2/3} \sqrt[3]{\omega }}\right)^2}{\kappa ^{4/3} \omega ^{5/3}}~+~\frac{\omega  \text{Ai}\left(\frac{\omega ~+~\omega '}{\kappa ^{2/3} \sqrt[3]{\omega '}}\right)^2}{\kappa ^{4/3} \omega '^{5/3}}.\label{betaT1}
\end{equation}
Considering the close association between accelerating mirrors and charges, we postulate that the electron spectrum has the same form as Equation~(\ref{betaT1}). See Figure \ref{contour} for an illustration of the symmetry between the modes, and see Figure \ref{Number} for a plot of the spectrum after integration over $\omega'$,
\begin{equation}
 N_\omega = \int_0^{\infty}  |\beta_{\omega\omega'}|^2 d \omega'. \label{Nw}
 \end{equation} 

\unskip

The reason for the agreement between emitted energy is related to the Lorentz symmetry group of Maxwell's equations in classical electrodynamics and the M\"obius symmetry group PSL(2,R) of the quantum moving mirror model. The quotient PSL(2,R) group is the restricted Lorentz group of 3~+~1-dimensional Minkowski space.  The moving mirror model can be seen to obey M\"obius symmetry, most explicitly by examining the Schwarzian derivative for the energy flux emission~\cite{Birrell:1982ix}, although the particle count and entanglement entropy are invariant under M\"obius transforms of the trajectory as well~\cite{mobmir}. 

From a scaling perspective, the classical electron Larmor power has exactly the same scaling on proper acceleration as the quantum mirror power. The same is true for the self-force of both the electron and mirror. 

The analog method works in part because of Lorentz invariance being the main underlying postulate of symmetry. The invariance manifests itself in analogies relating to mirrors, electrons and black holes with respect to both quantum and classical phenomena~\cite{Zhakenuly:2021pfm}.  For a mirror moving in 3~+~1D, the radiation obeys the quantum power formula~\cite{Good:2021ffo} 
\begin{equation}
P_{\textrm{mirror}} = \frac{\hbar \alpha^2}{6\pi c^2} \to \frac{\alpha^2}{6\pi}\,, 
\end{equation}
where $\alpha$ is the frame-invariant proper acceleration. Independent of dimension, i.e., the speed of light, this measure of power is Lorentz invariant. 
This can be compared to the classical counterpart, as we have seen, with the same relativistic covariant scaling that is well-studied in graduate classical electrodynamics~\cite{Jackson:490457} of the Larmor formula,
\begin{equation}P_{\textrm{electron}}= \frac{2}{3}\frac{q^2\alpha^2}{4\pi \epsilon_0 c^3} \to \frac{q^2\alpha^2}{6\pi},\end{equation} 
where the left side is in SI units, ``$\to$'' implies conversion to natural units, where $\epsilon_0 =c=\hbar = 1$, and $\alpha$ is the magnitude of the proper acceleration of the moving point charge. 

The Lorentz invariance cross-over is present in the quantum Lorentz--Abraham--Dirac (LAD) force discovered by Ford and Vilenkin~\cite{Ford:1982ct} (for a derivation see~\cite{Myrzakul:2021bgj}), the magnitude of which is a Lorentz scalar invariant jerk
\begin{equation}
F_{\textrm{mirror}} = \frac{\hbar \alpha'}{6\pi c^2} \to \frac{\alpha'}{6\pi}, \label{self-force}
\end{equation}
where the prime denotes a derivative with respect to proper time.  For convenience, this is compared to the classical LAD self-force with scalar magnitude, as we have seen: 
\begin{equation}
F_{\textrm{electron}} = \frac{q^2 \alpha'}{6\pi \epsilon_0 c^3} \to \frac{q^2 \alpha'}{6\pi}. \label{self-force}
\end{equation}
These are quantum and classical forces (and power), tying together mirror and electrons in close analogy. The moving mirror system, as an analog, thus serves to provide a new way of looking at the electron acceleration radiation problem, permitting novel application of the ideas in the quantum system to the corresponding classical system.

\section{Consistency checks}
Using the retarded time clock of the observer, the following integration holds:
\begin{equation}
E= \int_{-\infty}^{\infty} \bar{P}(u) d u = -\int_{-\infty}^{\infty} \bar{F}(u) d u , \label{barPF}
\end{equation}
demonstrating that both the Larmor and Feynman powers from Equations~(\ref{barP}) and~(\ref{barF}) are consistent with the conservation of globally radiated energy.  The total radiated energy from the Larmor power equals the total energy loss from the radiation reaction, as provided by the total energy associated with the Feynman power.   Despite $\bar{F} \neq \bar{P}$, there is no discrepancy between the total mechanical energy lost and the total radiated energy lost, in full agreement with the conclusions of Singal~\cite{Singal:2020kan,Singal_2018,Singal:2014fha}.  The Bogoliubov spectrum from Equation~(\ref{betaT1}) is consistent with the above results as well, because 
\begin{equation} 
E = \int_0^\infty \int_0^\infty \omega |\beta_{\omega\omega'}|^2 d \omega d \omega'. \label{numerical1}
\end{equation}
We may also sum $\omega'$ quanta by replacing the above integrand with $\omega' |\beta_{\omega\omega'}|^2$ and, by mode symmetry, obtaining the total radiated energy  $E$.
The quanta summing in Equation~(\ref{numerical1}) demonstrates that the radiated particles carry the energy. Thus, the total emitted energy is~analytic:
\begin{equation}
E = \frac{5\kappa}{96}.\label{energy}
\end{equation}\\
This establishes that the total emitted energy of the radiation is finite, positive, and dependent on detector scale sensitivity range set by the acceleration scale $\kappa$. Despite the asymptotic approach to the speed of light and the ever-increasing infinite energy used to accelerate the massive point charge, communication is severed between the acceleration energy and the radiation energy. The electron does not explode. 

\section{Discussion}
\begin{figure}[H]
  \includegraphics[width=0.95\linewidth]{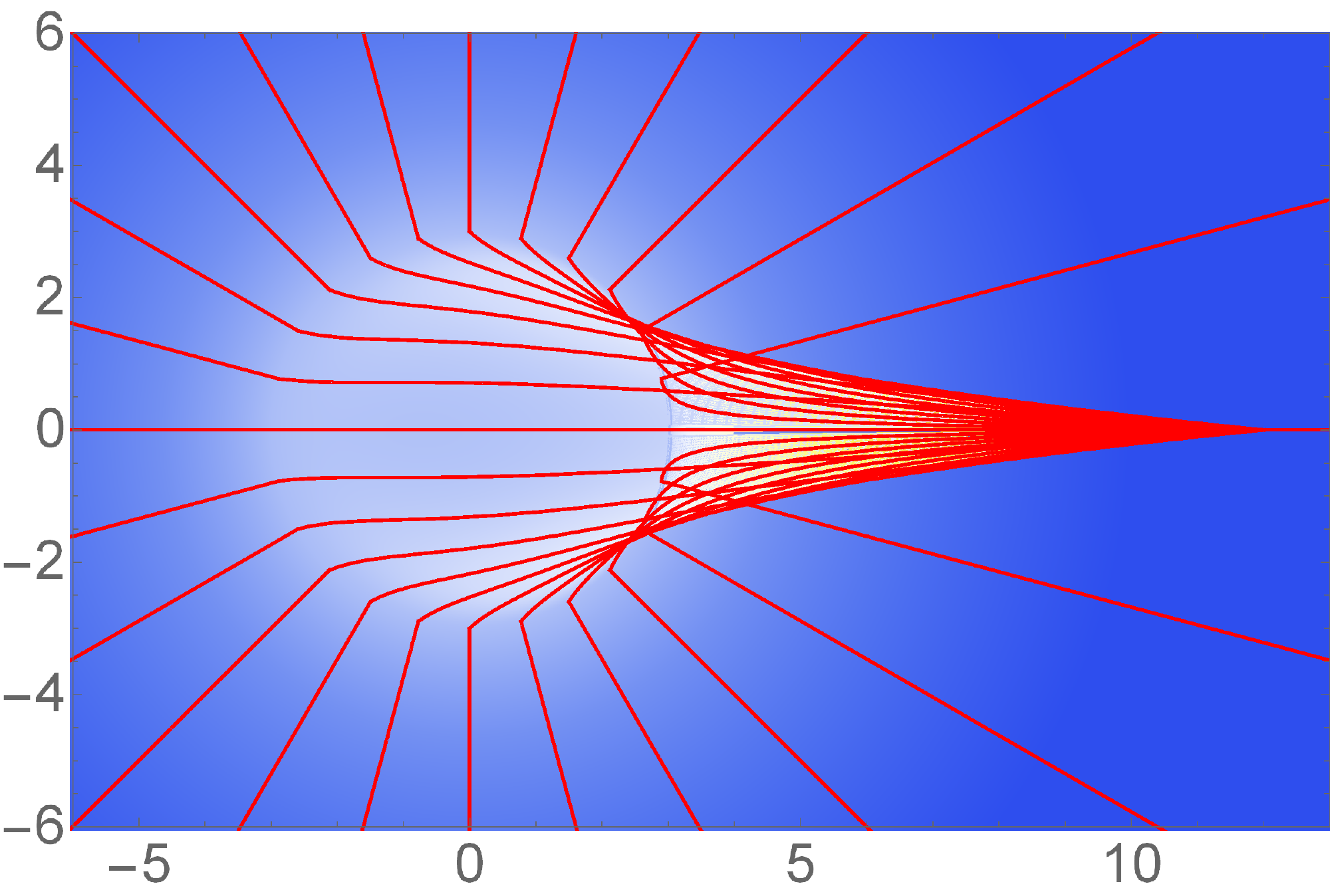}
 \caption{Electric field (density plot) and electric field lines
(red) for the accelerated point charge of Equation~(\ref{p(u)}) with $\kappa = 1$.  Here, $p(u) = 0$ before $u<0$ in order to reproduce the field lines of a static charge at large distances. The lines of the non-uniformly moving charge at small distances from the particle are tightly stretched ($u=3$).  The acceleration-related
transition zone in between is outpaced by the charge itself. The  numerical calculations for this figure were performed with the Ruhlandt--M\"uhle--Enderlein code~\cite{lines}.}
\label{lines}
\end{figure}

Kinetic processes responsible for converting the outflow energy of blazars, pulsar wind nebulae, and gamma ray bursters into electron acceleration radiation allow electrons to be accelerated to very high Lorentz factors before radiation damping is appreciable; see, e.g.,~\cite{Liang:2009cr}.  The electric fields lines experience a tight stretching effect due to their non-uniform acceleration relative to uniform acceleration (see Figures 2 and 3 of~\cite{lines}).  An illustration of the electric field lines is shown in Figure \ref{lines}.

The Light and Airy trajectory from Equation~(\ref{p(u)}) is the extreme case of `high Lorentz factor', where the kinetic energy completely outpaces the radiation energy.  The acceleration radiation experiences an asymptotic stop to further outflow. Using the scale set by IB beta decay in~\cite{Good:2022eub}, where $\kappa = 48 \alpha c \omega_\gamma$, $\omega_\gamma = \omega_\textrm{max} - \omega_\textrm{min}$ and $E_\gamma = \hbar \omega_\gamma$, we can convert Equation~(\ref{energy}) to SI units, resulting in  $E = 5\alpha E_\gamma/2$, where $\alpha$ is the fine structure constant.  Thus, in this scale the energy emitted is dependent on the  sensitivity range of the detector, suggesting independence from the responsible kinetic processes.  This makes sense, as conservation of energy is not in line with the total divergent kinetic energy, at least in the sense that the frequencies for appreciable radiation should be limited to the total energy $E_T$ provided to the electron, $\omega_{\textrm{max}} < E_T/\hbar$.  

The Light and Airy trajectory works for finite radiation energy because of strong asymptotic inertia coupled to motion near the speed of light; interestingly, the reality of the beta Bogoliubov coefficient~\cite{Good:2021dkh} leads to fixed radiation that is analytic and simple.  

The simplest examples of non-trivial beta Bogoliubov transformations are those in which the Bogoliubov coefficients are real-valued. The simplicity of a real non-zero beta Bogoliubov coefficient for particle creation is likely connected to the simplicity of the associated trajectory and finite radiation energy. 

To guarantee a real-valued beta Bogoliubov coefficient, the trajectory $v(u)$ must be an odd function to ensure that the exponential over the symmetric interval for the beta coefficient turns into a cosine of the argument, i.e., a real valued function. The simplest odd function that accelerates with asymptotic inertia and accelerates to the speed of light is $v(u) \sim u~+~u^3$ (Equation~(\ref{p(u)})).

\section{Conclusion}
In this note, we have avoided the radiation catastrophe that accompanies asymptotic light speed.  With strong asymptotic inertia, our proposed solution demonstrates that infinite energy radiance need not occur in conjunction with asymptotic light speed approach.  Despite the nonphysical requirement of infinite kinetic energy necessary to push an electron along the proposed motion, the total radiated energy is fixed and physically consistent with the radiation reaction, power emission, and acceleration. 

The trajectory demonstrates the physical separation and interplay between kinetic energy loss and radiative energy loss. Moreover, the motion demonstrates how to avoid the formation of a horizon despite the advance toward the speed of light.  Mapping to curved spacetime and finding the analog geometry would be an interesting topic for future study.

\section*{Acknowledgments}
Funding comes in part from the FY2021-SGP-1-STMM Faculty Development Competitive Research Grant No. 021220FD3951 at Nazarbayev University.
\bibliography{main} 
\end{document}